\documentclass[11pt]{article}

\usepackage[preprint]{acl}

\usepackage{times}
\usepackage{latexsym}

\usepackage[T1]{fontenc}

\usepackage[utf8]{inputenc}

\usepackage{microtype}
\usepackage{url}
\usepackage{enumitem}
\usepackage{booktabs}
\usepackage{arydshln}
\usepackage{tcolorbox}
\tcbuselibrary{breakable}
\usepackage{amssymb}

\usepackage{inconsolata}

\usepackage{graphicx}
\usepackage{hyperref}
\usepackage{multirow}

\title{Dual-View Training for Instruction-Following Information Retrieval}

\author{
Qingcheng Zeng\textsuperscript{$\dagger, $}\thanks{Work done during internship at Snowflake Inc.},
Puxuan Yu\textsuperscript{$\dagger$},
Aman Mehta\textsuperscript{$\dagger$},
Fuheng Zhao\textsuperscript{$\dagger$},
Rajhans Samdani\textsuperscript{$\dagger$}\\\\
\textsuperscript{$\ast$}Northwestern University
\textsuperscript{$\dagger$}Snowflake Inc.
}

\begin{document}
\maketitle
\begin{abstract}
Instruction-following information retrieval (IF-IR) studies retrieval systems that must not only find documents relevant to a query, but also obey explicit user constraints such as required attributes, exclusions, or output preferences. However, most retrievers are trained primarily for semantic relevance and often fail to distinguish documents that match the topic from those that satisfy the instruction. We propose a dual-view data synthesis strategy based on polarity reversal: given a query, a document that is relevant under the instruction, and a hard negative that matches the query but violates the instruction, we prompt an LLM to generate a complementary instruction under which the two documents swap relevance labels. By presenting the same document pair under complementary instructions that invert their relevance labels, the training signal forces the retriever to reconsider the same candidate set through the instruction, rather than relying on fixed topical cues. On a 305M-parameter encoder, our method improves performance on the FollowIR benchmark by 45\%, surpassing general-purpose embedding models of comparable or larger scale. Through head-to-head comparisons at matched data budgets, we further show that data diversity and instruction supervision play complementary roles: the former preserves general retrieval quality, while the latter improves instruction sensitivity. These results highlight the value of targeted data synthesis for building retrieval systems that are both broadly capable and instruction-aware.
\end{abstract}

\section{Introduction}
Instruction-following information retrieval extends traditional semantic matching by requiring systems to adhere to both a query and explicit user-defined constraints that specify relevance criteria \cite{su-etal-2023-one,weller-etal-2025-followir}. For instance, a user might not only submit a query but also specify that relevant documents must discuss a particular aspect, be written in a certain style, or satisfy length requirements. This paradigm rigorously tests the capacity of dense retrievers to adapt their behavior based on dynamic in-context directives, going beyond static notions of relevance.

Despite the growing number of instruction-aware retrievers, critical limitations persist. \citet{weller-etal-2025-followir} conducted a systematic evaluation using human-annotated instructions that fundamentally alter relevance definitions. By quantifying sensitivity to instruction changes with the $p$-MRR metric, which measures whether a retriever ranks the preferred document higher when the instruction changes, their findings reveal that most current models fail to internalize detailed relevance criteria, relying instead on superficial query-document similarity and largely ignoring the specific constraints imposed by instructions.

\begin{figure}
    \centering
    \includegraphics[width=\columnwidth]{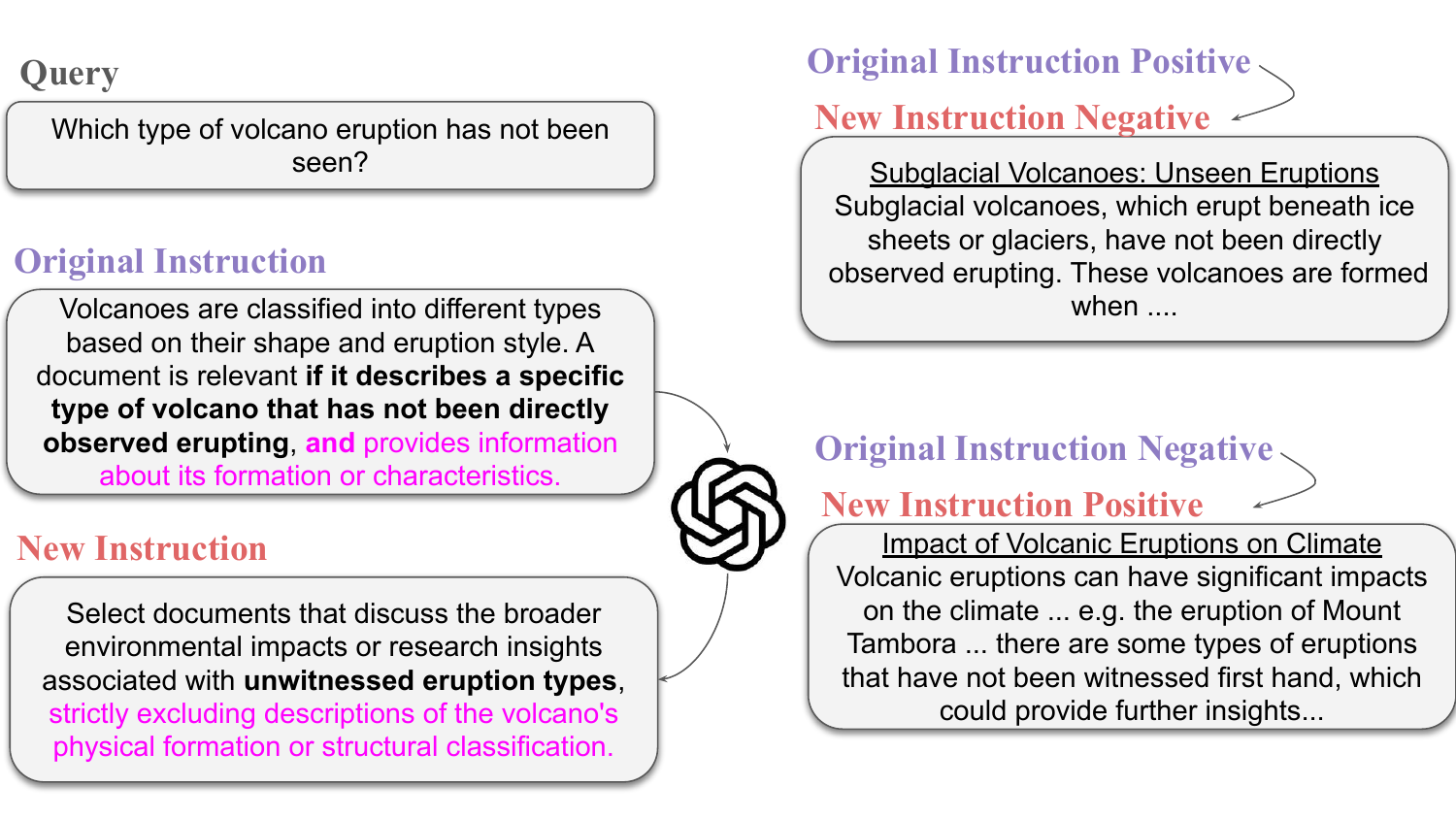}
    \caption{We synthesize new instructions that reverse the relevance polarity of existing documents, creating challenging samples that sharpen the retriever's sensitivity to instructional nuances.}
    \label{fig:teaser}
\end{figure}

To address this, \citet{ICLR2025_promptriever} introduced a training paradigm centered on \textit{instruction negatives}, documents that are semantically relevant to the query but become irrelevant once a specific instruction is applied. While their results demonstrate the effectiveness of instruction negatives over standard hard negatives, these negatives carry additional untapped potential: each one implicitly defines a complementary instruction under which it becomes the relevant document. In this paper, we exploit this observation by prompting an LLM to synthesize such complementary instructions that reverse the relevance polarity of existing document pairs (\autoref{fig:teaser}). The same documents are thus repurposed under two complementary views, compelling the retriever to attend to fine-grained instructional differences rather than surface-level query-document similarity.

We make three contributions. (1) We propose a simple polarity-reversal synthesis strategy that improves FollowIR $p$-MRR by 45\% on a 305M-parameter encoder, surpassing general-purpose embedding models of comparable or larger scale. (2) Through head-to-head comparisons at matched data budgets, we identify a fundamental tension in IF training: data diversity sustains general retrieval quality while instruction supervision drives IF capability, but supplementing with non-instruct data may dilute the instruction signal and hurt IF performance. (3) We show that dual-view synthesis resolves this tension, simultaneously improving instruction sensitivity and general retrieval at equal data budget. All findings are validated across two encoder backbones.

\section{Methodology}

As illustrated in \autoref{fig:teaser}, our approach leverages LLMs to generate complementary instructions that invert the ground truth labels for a fixed set of documents. Formally, given a query $q$, a positive document $D^+$, and an instruction negative document $D^-$ under an original instruction $I_{orig}$, we prompt the LLM to synthesize a new instruction $I_{new}$. The generation is constrained such that $I_{new}$ must be semantically coherent with $q$ but sufficiently distinct from $I_{orig}$ so that $D^-$ becomes the relevant document (positive) and $D^+$ becomes an instruction negative. This creates a ``dual-view'' training scenario where the relevance of a document depends entirely on the specific constraints of the instruction, not on its text or the query alone. By requiring the model to retrieve opposite documents for the same query under different instructions, this setup discourages reliance on fixed query-document associations and encourages generalization across diverse instruction types.

During contrastive training, each query is paired with its positive document and a set of negatives. For datapoints augmented with our method, the training batch contains both the original and the polarity-reversed view. The model must therefore learn to assign high similarity to $(q \oplus I_{orig}, D^+)$ and $(q \oplus I_{new}, D^-)$ simultaneously, while pushing the reversed assignments apart. This dual objective directly penalizes instruction-agnostic representations: no single query encoding can satisfy both constraints unless it genuinely conditions on the semantic content of the instruction, since $q \oplus I_{orig}$ and $q \oplus I_{new}$ must retrieve \textit{opposite} documents.

\paragraph{Data Synthesis Setup} We employ \textit{Qwen3-Next-80B-A3B-Instruct} \cite{qwen3technicalreport} as the backbone LLM for our data synthesis pipeline. We construct our seed dataset by selecting instances from the promptriever data that contain at least one pre-existing instruction negative. For each of these data points, we generate a new instruction that reverses the roles of the positive document and the instruction negative. This process yields one complementary training sample per original instance. In our controlled experiments, DV samples substitute for an equal-sized portion of the training set rather than being added on top, enabling size-matched comparisons across all configurations. After synthesis, one annotator manually checked 100 datapoints and confirmed that over 99\% of the DV instructions are usable. Thus, no additional filtering was conducted. The specific prompt template used for this generation is provided in \autoref{appendix:prompt_template}.

\section{Experimental Setup}

\paragraph{Backbone Models} We adopt \textit{gte-multilingual-mlm-base} \cite{zhang-etal-2024-mgte} (305M parameters) as our primary encoder, initialized from our own contrastively pretrained checkpoint trained on 1.41 billion unsupervised query-document pairs from C4 \cite{JMLR:v21:20-074}, mC4 \cite{habernal-etal-2016-c4corpus}, CC News, and multilingual Wikipedia. To assess cross-backbone generalizability, we additionally train \textit{bge-m3-retromae} \cite{chen-etal-2024-m3} under the same data configurations and report results in \autoref{tab:results_of_bge-m3-retromae}.

\begin{itemize}[itemsep=0pt, topsep=2pt, parsep=0pt, partopsep=0pt]
  \item \textit{Ins-orig}: 480k instruct samples from the original Promptriever dataset.
  \item \textit{Ins-DV} (ours): 240k original instruct samples + 240k dual-view synthetic samples (size-matched to Ins-orig).
  \item \textit{All-orig}: 440k instruct samples + their 440k non-instruct counterparts from the original Promptriever dataset (each instruct sample was synthesized from a corresponding non-instruct sample).
  \item \textit{All-DV} (ours): 440k original instruct samples + 440k dual-view synthetic samples (size-matched to All-orig).
\end{itemize}

\paragraph{Training Details} We use the Arctic-Embed framework\footnote{\url{https://github.com/snowflakedb/ArcticTraining}} with 30 hard negatives per query, including 1--3 instruction negatives. We optimize with InfoNCE loss \cite{oord2019representationlearningcontrastivepredictive} with temperature $\tau = 0.02$. For both encoders, the query and instruction are concatenated before encoding, while documents are encoded independently. The maximum sequence length is 512 tokens for both queries and documents. All configurations use the same training protocol to ensure fair comparisons.

\paragraph{Evaluation Benchmarks} We evaluate on three benchmark suites: (1) \textbf{FollowIR} \cite{weller-etal-2025-followir}, reporting $p$-MRR for instruction sensitivity and an aggregated Score for overall retrieval quality. The $p$-MRR metric quantifies instruction sensitivity by comparing a model's ranking under two paired instructions: one where a document is annotated as relevant and one where it is not. Positive $p$-MRR indicates the model correctly adjusts its ranking in response to instruction changes, while negative values indicate it ranks documents in the \textit{opposite} direction. (2) \textbf{InfoSearch} \cite{infosearch} \textit{length} and \textit{keyword} subsets, also reporting $p$-MRR; and (3) \textbf{MAIR} \cite{sun-etal-2024-mair} IFEval \cite{zhou2023instructionfollowingevaluationlargelanguage} and InstructIR \cite{oh2024instructirbenchmarkinstructionfollowing} subsets, reporting nDCG@10.

\section{Results and Analysis}

\begin{table*}[t]
\centering
\small
\begin{tabular}{lcccccc}
\toprule
\multirow{2}{*}{Training data} &
\multicolumn{2}{c}{FollowIR} &
\multicolumn{2}{c}{InfoSearch (p-MRR)} &
\multicolumn{2}{c}{MAIR (nDCG@10)} \\
\cmidrule(lr){2-3}\cmidrule(lr){4-5}\cmidrule(lr){6-7}
& p-MRR $\uparrow$ & Score $\uparrow$ & Length $\uparrow$ & Keyword $\uparrow$
& IFEval $\uparrow$ & InstructIR $\uparrow$ \\
\midrule
Ins-orig        & 5.21 & 21.33 & 4.06 & 2.06 & 32.14 & 89.16 \\
Ins-DV (ours)  & 7.57 & 19.73 & 9.02 & 5.61 & 36.13 & 87.97 \\
\cdashline{1-7}
All-orig        & 5.27 & 20.85 & -23.22 & -49.65 & 24.33 & 85.54 \\
All-DV (ours)  & 8.30 & 21.38 & 31.91 & 12.13 & 34.08 & 90.74 \\
\bottomrule
\end{tabular}
\caption{Main results on instruction-following retrieval benchmarks. Ins-/All- denote instruct-only and mixed training regimes; -orig/-DV denote original and our dual-view augmented data. Score is the macro-average across three FollowIR subsets (MAP@1000 on two subsets; nDCG@5 on one). Negative $p$-MRR indicates the model \textit{contradicts} instruction-defined relevance.}
\label{tab:main_results}
\end{table*}

\begin{table*}[t]
\centering
\small
\begin{tabular}{lcccccc}
\toprule
\multirow{2}{*}{Training data} &
\multicolumn{2}{c}{FollowIR} &
\multicolumn{2}{c}{InfoSearch (p-MRR)} &
\multicolumn{2}{c}{MAIR (nDCG@10)} \\
\cmidrule(lr){2-3}\cmidrule(lr){4-5}\cmidrule(lr){6-7}
& p-MRR $\uparrow$ & Score $\uparrow$ & Length $\uparrow$ & Keyword $\uparrow$
& IFEval $\uparrow$ & InstructIR $\uparrow$ \\
\midrule
Ins-orig        & 9.40  & 22.26 & 19.00  & 4.18   & 27.12 & 90.40 \\
Ins-DV (ours)  & 11.47 & 19.76 & 28.64  & 48.42  & 29.37 & 89.53 \\
\cdashline{1-7}
All-orig        & 8.84  & 20.69 & -27.08 & -62.04 & 25.68 & 90.67 \\
All-DV (ours)  & 13.92 & 20.99 & 40.15  & 49.62  & 27.07 & 90.80 \\
\bottomrule
\end{tabular}
\caption{Results on \textit{bge-m3-retromae}. Score is computed identically to \autoref{tab:main_results}. The same patterns hold: DV data improves IF, mixing non-instruct data degrades it, and DV augmentation counteracts this degradation.}
\label{tab:results_of_bge-m3-retromae}
\end{table*}

\autoref{tab:main_results} summarizes results across all benchmarks. We organize our analysis around two experimental comparisons that together reveal the interplay between instruction sensitivity and data diversity.

\paragraph{Instruct-only comparison: IF gains at the cost of general retrieval.}
In the size-matched comparison between Ins-DV and Ins-orig ($\sim$480k samples each), our DV data yields consistent gains across all IF metrics: FollowIR $p$-MRR increases from 5.21 to 7.57 (+45\%), surpassing general-purpose models such as EmbeddingGemma-300M (5.61 $p$-MRR) \cite{vera2025embeddinggemmapowerfullightweighttext} and Qwen3-Embedding-0.6B (5.09 $p$-MRR) \cite{zhang2025qwen3embeddingadvancingtext}; InfoSearch $p$-MRR improves by +122\% (length) and +172\% (keyword); and MAIR IFEval rises from 32.14 to 36.13. However, FollowIR Score drops from 21.33 to 19.73. Since Ins-DV replaces half of the original instruct samples with DV counterparts, the model sees fewer unique training contexts, which may account for this decline. This observation points to data diversity as a key factor in sustaining general retrieval performance, a hypothesis we test directly in the All- configurations below.

\paragraph{Mixed-data comparison: All-orig vs.\ All-syn.}
The All- configurations compare two strategies for scaling the training set to $\sim$880k while holding the instruct portion fixed ($\sim$440k): supplementing with non-instruct data (All-orig) versus DV data (All-DV). Across all benchmarks, All-DV outperforms All-orig. FollowIR $p$-MRR improves from 5.27 to 8.30, the highest among all configurations, and Score rises from 20.85 to 21.38. On InfoSearch, All-DV achieves positive $p$-MRR (31.91 length, 12.13 keyword) whereas All-orig falls into negative territory ($-23.22$, $-49.65$), indicating that the model with non-instruct supplementation contradicts instruction-defined relevance. MAIR metrics follow the same pattern: All-DV reaches 34.08 (IFEval) and 90.74 (InstructIR), both the best across all settings.

\paragraph{The role of data diversity.}
These results, combined with the Ins- comparison, clarify the respective roles of data diversity and instruction supervision. The Ins- experiments show that dedicated instruction data drives IF capability, but replacing original samples with DV ones reduces diversity and costs general retrieval quality. The All- experiments reveal the converse: scaling with non-instruct data provides diversity but dilutes the instruction signal, severely degrading IF while offering only marginal general retrieval benefit over Ins-orig. All-DV achieves the best of both by providing instruction-conditioned training pairs at scale, simultaneously maintaining the data volume that sustains general quality and the instruction signal that drives IF capability. Notably, All-DV contains no non-instruct data, yet achieves the best Score across all configurations. This suggests that data volume, rather than source heterogeneity per se, is the primary driver of general retrieval quality, as long as individual training samples are sufficiently diverse in their query-document pairings.

\paragraph{Cross-backbone generalizability.}
\autoref{tab:results_of_bge-m3-retromae} reports results on \textit{bge-m3-retromae}, a stronger encoder with a different pretraining strategy. Both experimental patterns replicate faithfully. In the Ins- comparison, Ins-DV improves all IF metrics: FollowIR $p$-MRR rises from 9.40 to 11.47 (+22\%), InfoSearch keyword $p$-MRR surges from 4.18 to 48.42, while Score drops from 22.26 to 19.76, confirming the same IF/diversity trade-off. In the All- comparison, the same degradation pattern reappears with All-orig ($-27.08$ and $-62.04$ on InfoSearch), and All-DV again reverses it entirely (40.15 and 49.62), achieving the best FollowIR $p$-MRR of 13.92 while maintaining competitive Score (20.99). The magnitude of the keyword gains is considerably larger on bge-m3, suggesting that a stronger backbone amplifies the benefit of our DV signal. The cross-model consistency confirms that both the DV method and the data mixing degradation are backbone-agnostic phenomena, reinforcing the generality of our findings.

\section{Discussion}

\paragraph{Polarity reversal versus instruction-based negatives.}
Both Promptriever \cite{ICLR2025_promptriever} and InF-IR \cite{zhuang2025infir} demonstrate that instruction-tied negatives outperform generic hard negatives, but treat negatives as fixed failures relative to a given instruction, i.e., documents that should not be retrieved under it. Polarity reversal reframes this: an instruction negative is a \textit{conditionally relevant} document, one that should be retrieved under a different, complementary instruction. Synthesizing this complement imposes a contrastive constraint \textit{across} instruction space, requiring the query encoder to resolve where $I_{orig}$ and $I_{new}$ diverge, not just which documents each instruction excludes. This targets instructional distinctions rather than instructional exclusions, a structurally richer supervisory signal.

\paragraph{A gradient perspective on data mixing.}
The data mixing catastrophe offers a mechanistic explanation for why general-purpose embedding models often underperform instruction-specialized ones despite larger scale. Non-instruct samples provide gradient signal that rewards query-correlated retrieval regardless of instructions; at a 50/50 mix, this overwhelms the instruction signal. Instruction sensitivity is therefore not a capability that accumulates with scale but a fragile property requiring \textit{consistent} supervision. This is consistent with InF-IR \cite{zhuang2025infir}, which achieves competitive IF performance from $\sim$38k specialized triplets, suggesting signal purity matters more than volume. Our DV strategy addresses this by embedding an instruction-conditioning signal into every training pair, providing a uniform gradient toward instruction-conditioned representations even in the presence of general-retrieval data.

\section{Conclusion}
We present a dual-view data synthesis strategy based on polarity reversal that creates complementary training pairs at no additional annotation cost. Experiments across two encoder backbones yield two insights: (1) dedicated instruction data drives IF sensitivity, while data diversity sustains general retrieval quality, and (2) our synthesis reconciles these competing demands, simultaneously improving both dimensions at equal data budget. The approach requires no changes to existing pipelines.

\section*{Limitations}
The polarity-reversal synthesis assumes that a meaningful complementary instruction exists for each data point; in practice, our manual inspection found this to hold for the vast majority of cases, but queries with very narrow relevance criteria may occasionally yield less natural reversals. We evaluate on encoder-based bi-encoder retrievers; exploring the applicability to decoder-based or cross-encoder architectures is a natural direction for future work. Additionally, our experiments focus on English-language benchmarks, and extending the approach to multilingual settings remains an interesting avenue to explore.

\bibliography{custom}

@inproceedings{su-etal-2023-one,
    title = "One Embedder, Any Task: Instruction-Finetuned Text Embeddings",
    author = "Su, Hongjin  and
      Shi, Weijia  and
      Kasai, Jungo  and
      Wang, Yizhong  and
      Hu, Yushi  and
      Ostendorf, Mari  and
      Yih, Wen-tau  and
      Smith, Noah A.  and
      Zettlemoyer, Luke  and
      Yu, Tao",
    editor = "Rogers, Anna  and
      Boyd-Graber, Jordan  and
      Okazaki, Naoaki",
    booktitle = "Findings of the Association for Computational Linguistics: ACL 2023",
    month = jul,
    year = "2023",
    address = "Toronto, Canada",
    publisher = "Association for Computational Linguistics",
    url = "https://aclanthology.org/2023.findings-acl.71/",
    doi = "10.18653/v1/2023.findings-acl.71",
    pages = "1102--1121"
}

@inproceedings{weller-etal-2025-followir,
    title = "{F}ollow{IR}: Evaluating and Teaching Information Retrieval Models to Follow Instructions",
    author = "Weller, Orion  and
      Chang, Benjamin  and
      MacAvaney, Sean  and
      Lo, Kyle  and
      Cohan, Arman  and
      Van Durme, Benjamin  and
      Lawrie, Dawn  and
      Soldaini, Luca",
    editor = "Chiruzzo, Luis  and
      Ritter, Alan  and
      Wang, Lu",
    booktitle = "Proceedings of the 2025 Conference of the Nations of the Americas Chapter of the Association for Computational Linguistics: Human Language Technologies (Volume 1: Long Papers)",
    month = apr,
    year = "2025",
    address = "Albuquerque, New Mexico",
    publisher = "Association for Computational Linguistics",
    url = "https://aclanthology.org/2025.naacl-long.597/",
    doi = "10.18653/v1/2025.naacl-long.597",
    pages = "11926--11942",
    ISBN = "979-8-89176-189-6"
}

@inproceedings{ICLR2025_promptriever,
 author = {Weller, Orion and Van Durme, Ben and Lawrie, Dawn and Paranjape, Ashwin and Zhang, Yuhao and Hessel, Jack},
 booktitle = {International Conference on Representation Learning},
 editor = {Y. Yue and A. Garg and N. Peng and F. Sha and R. Yu},
 pages = {17660--17683},
 title = {Promptriever: Instruction-Trained Retrievers Can Be Prompted Like Language Models},
 url = {https://proceedings.iclr.cc/paper_files/paper/2025/file/2cefdb2c4c3274b78cd450bac35228df-Paper-Conference.pdf},
 volume = {2025},
 year = {2025}
}

@misc{qwen3technicalreport,
      title={Qwen3 Technical Report}, 
      author={{Qwen Team}},
      year={2025},
      eprint={2505.09388},
      archivePrefix={arXiv},
      primaryClass={cs.CL},
      url={https://arxiv.org/abs/2505.09388}, 
}

@inproceedings{zhang-etal-2024-mgte,
    title = "{mGTE}: Generalized Long-Context Text Representation and Reranking Models for Multilingual Text Retrieval",
    author = "Zhang, Xin  and
      Zhang, Yanzhao  and
      Long, Dingkun  and
      Xie, Wen  and
      Dai, Ziqi  and
      Tang, Jialong  and
      Lin, Huan  and
      Yang, Baosong  and
      Xie, Pengjun  and
      Huang, Fei  and
      Zhang, Meishan  and
      Li, Wenjie  and
      Zhang, Min",
    editor = "Dernoncourt, Franck  and
      Preo{\c{t}}iuc-Pietro, Daniel  and
      Shimorina, Anastasia",
    booktitle = "Proceedings of the 2024 Conference on Empirical Methods in Natural Language Processing: Industry Track",
    month = nov,
    year = "2024",
    address = "Miami, Florida, US",
    publisher = "Association for Computational Linguistics",
    url = "https://aclanthology.org/2024.emnlp-industry.103/",
    doi = "10.18653/v1/2024.emnlp-industry.103",
    pages = "1393--1412"
}

@inproceedings{chen-etal-2024-m3,
    title = "{M}3-Embedding: Multi-Linguality, Multi-Functionality, Multi-Granularity Text Embeddings Through Self-Knowledge Distillation",
    author = "Chen, Jianlyu  and
      Xiao, Shitao  and
      Zhang, Peitian  and
      Luo, Kun  and
      Lian, Defu  and
      Liu, Zheng",
    editor = "Ku, Lun-Wei  and
      Martins, Andre  and
      Srikumar, Vivek",
    booktitle = "Findings of the Association for Computational Linguistics: ACL 2024",
    month = aug,
    year = "2024",
    address = "Bangkok, Thailand",
    publisher = "Association for Computational Linguistics",
    url = "https://aclanthology.org/2024.findings-acl.137/",
    doi = "10.18653/v1/2024.findings-acl.137",
    pages = "2318--2335"
}

@inproceedings{habernal-etal-2016-c4corpus,
    title = "{C}4{C}orpus: Multilingual Web-size Corpus with Free License",
    author = "Habernal, Ivan  and
      Zayed, Omnia  and
      Gurevych, Iryna",
    editor = "Calzolari, Nicoletta  and
      Choukri, Khalid  and
      Declerck, Thierry  and
      Goggi, Sara  and
      Grobelnik, Marko  and
      Maegaard, Bente  and
      Mariani, Joseph  and
      Mazo, Helene  and
      Moreno, Asuncion  and
      Odijk, Jan  and
      Piperidis, Stelios",
    booktitle = "Proceedings of the Tenth International Conference on Language Resources and Evaluation ({LREC}'16)",
    month = may,
    year = "2016",
    address = "Portoro{\v{z}}, Slovenia",
    publisher = "European Language Resources Association (ELRA)",
    url = "https://aclanthology.org/L16-1146/",
    pages = "914--922"
}

@article{JMLR:v21:20-074,
  author  = {Colin Raffel and Noam Shazeer and Adam Roberts and Katherine Lee and Sharan Narang and Michael Matena and Yanqi Zhou and Wei Li and Peter J. Liu},
  title   = {Exploring the Limits of Transfer Learning with a Unified Text-to-Text Transformer},
  journal = {Journal of Machine Learning Research},
  year    = {2020},
  volume  = {21},
  number  = {140},
  pages   = {1--67},
  url     = {http://jmlr.org/papers/v21/20-074.html}
}

@misc{oord2019representationlearningcontrastivepredictive,
      title={Representation Learning with Contrastive Predictive Coding}, 
      author={Aaron van den Oord and Yazhe Li and Oriol Vinyals},
      year={2019},
      eprint={1807.03748},
      archivePrefix={arXiv},
      primaryClass={cs.LG},
      url={https://arxiv.org/abs/1807.03748}, 
}

@inproceedings{infosearch,
 author = {Zhou, Jianqun and Zheng, Yuanlei and Chen, Wei and Zheng, Qianqian and Zeyuan, Shang and Zhang, Wei and Meng, Rui and Shen, Xiaoyu},
 booktitle = {International Conference on Representation Learning},
 editor = {Y. Yue and A. Garg and N. Peng and F. Sha and R. Yu},
 pages = {84965--84996},
 title = {Beyond Content Relevance: Evaluating Instruction Following in Retrieval Models},
 url = {https://proceedings.iclr.cc/paper_files/paper/2025/file/d37a4093931f359eb5fac5a25199db57-Paper-Conference.pdf},
 volume = {2025},
 year = {2025}
}

@inproceedings{sun-etal-2024-mair,
    title = "{MAIR}: A Massive Benchmark for Evaluating Instructed Retrieval",
    author = "Sun, Weiwei  and
      Shi, Zhengliang  and
      Long, Wu Jiu  and
      Yan, Lingyong  and
      Ma, Xinyu  and
      Liu, Yiding  and
      Cao, Min  and
      Yin, Dawei  and
      Ren, Zhaochun",
    editor = "Al-Onaizan, Yaser  and
      Bansal, Mohit  and
      Chen, Yun-Nung",
    booktitle = "Proceedings of the 2024 Conference on Empirical Methods in Natural Language Processing",
    month = nov,
    year = "2024",
    address = "Miami, Florida, USA",
    publisher = "Association for Computational Linguistics",
    url = "https://aclanthology.org/2024.emnlp-main.778/",
    doi = "10.18653/v1/2024.emnlp-main.778",
    pages = "14044--14067"
}

@misc{zhou2023instructionfollowingevaluationlargelanguage,
      title={Instruction-Following Evaluation for Large Language Models}, 
      author={Jeffrey Zhou and Tianjian Lu and Swaroop Mishra and Siddhartha Brahma and Sujoy Basu and Yi Luan and Denny Zhou and Le Hou},
      year={2023},
      eprint={2311.07911},
      archivePrefix={arXiv},
      primaryClass={cs.CL},
      url={https://arxiv.org/abs/2311.07911}, 
}

@misc{oh2024instructirbenchmarkinstructionfollowing,
      title={INSTRUCTIR: A Benchmark for Instruction Following of Information Retrieval Models}, 
      author={Hanseok Oh and Hyunji Lee and Seonghyeon Ye and Haebin Shin and Hansol Jang and Changwook Jun and Minjoon Seo},
      year={2024},
      eprint={2402.14334},
      archivePrefix={arXiv},
      primaryClass={cs.CL},
      url={https://arxiv.org/abs/2402.14334}, 
}

@misc{vera2025embeddinggemmapowerfullightweighttext,
      title={EmbeddingGemma: Powerful and Lightweight Text Representations}, 
      author={Henrique Schechter Vera and Sahil Dua and Biao Zhang and Daniel Salz and Ryan Mullins and Sindhu Raghuram Panyam and Sara Smoot and Iftekhar Naim and Joe Zou and Feiyang Chen and Daniel Cer and Alice Lisak and Min Choi and Lucas Gonzalez and Omar Sanseviero and Glenn Cameron and Ian Ballantyne and Kat Black and Kaifeng Chen and Weiyi Wang and Zhe Li and Gus Martins and Jinhyuk Lee and Mark Sherwood and Juyeong Ji and Renjie Wu and Jingxiao Zheng and Jyotinder Singh and Abheesht Sharma and Divyashree Sreepathihalli and Aashi Jain and Adham Elarabawy and AJ Co and Andreas Doumanoglou and Babak Samari and Ben Hora and Brian Potetz and Dahun Kim and Enrique Alfonseca and Fedor Moiseev and Feng Han and Frank Palma Gomez and Gustavo Hernández Ábrego and Hesen Zhang and Hui Hui and Jay Han and Karan Gill and Ke Chen and Koert Chen and Madhuri Shanbhogue and Michael Boratko and Paul Suganthan and Sai Meher Karthik Duddu and Sandeep Mariserla and Setareh Ariafar and Shanfeng Zhang and Shijie Zhang and Simon Baumgartner and Sonam Goenka and Steve Qiu and Tanmaya Dabral and Trevor Walker and Vikram Rao and Waleed Khawaja and Wenlei Zhou and Xiaoqi Ren and Ye Xia and Yichang Chen and Yi-Ting Chen and Zhe Dong and Zhongli Ding and Francesco Visin and Gaël Liu and Jiageng Zhang and Kathleen Kenealy and Michelle Casbon and Ravin Kumar and Thomas Mesnard and Zach Gleicher and Cormac Brick and Olivier Lacombe and Adam Roberts and Qin Yin and Yunhsuan Sung and Raphael Hoffmann and Tris Warkentin and Armand Joulin and Tom Duerig and Mojtaba Seyedhosseini},
      year={2025},
      eprint={2509.20354},
      archivePrefix={arXiv},
      primaryClass={cs.CL},
      url={https://arxiv.org/abs/2509.20354}, 
}

@misc{zhuang2025infir,
      title={Towards Better Instruction Following Retrieval Models},
      author={Yuchen Zhuang and Aaron Trinh and Rushi Qiang and Haotian Sun and Chao Zhang and Hanjun Dai and Bo Dai},
      year={2025},
      eprint={2505.21439},
      archivePrefix={arXiv},
      primaryClass={cs.IR},
      url={https://arxiv.org/abs/2505.21439},
}

@misc{zhang2025qwen3embeddingadvancingtext,
      title={Qwen3 Embedding: Advancing Text Embedding and Reranking Through Foundation Models}, 
      author={Yanzhao Zhang and Mingxin Li and Dingkun Long and Xin Zhang and Huan Lin and Baosong Yang and Pengjun Xie and An Yang and Dayiheng Liu and Junyang Lin and Fei Huang and Jingren Zhou},
      year={2025},
      eprint={2506.05176},
      archivePrefix={arXiv},
      primaryClass={cs.CL},
      url={https://arxiv.org/abs/2506.05176}, 
}

\appendix

\section{The Prompt Template for Data Synthesis}
\label{appendix:prompt_template}

The following box shows the full prompt template used for polarity-reversed instruction synthesis. Template variables (in \texttt{monospace}) are populated per instance.

\begin{tcolorbox}[colback=gray!3, colframe=gray!60, boxrule=0.5pt, arc=2pt, left=6pt, right=6pt, top=4pt, bottom=4pt, breakable]
\scriptsize

\textbf{Goal}\\
Create a new synthetic instruction that \textbf{reverses} the original relevance judgment \textbf{only} for the specified passages:
\begin{itemize}[nosep, leftmargin=*]
\item The \textbf{original positive passage (P$^{+}$)} must become an \textbf{instruction negative} under the new instruction (i.e., relevant to the pure query but \textbf{irrelevant} once the instruction is applied).
\item The \textbf{specific instruction negative (N$^{\star}$)} must become the \textbf{new positive} (i.e., relevant to the pure query \textbf{and} to the query+instruction).
\item \textbf{All remaining instruction negatives (N$_1$\ldots N$_k$)} must \textbf{remain instruction negatives}.
\end{itemize}
You must not change the query or any passage content. Only write a new instruction.

\medskip
\textbf{Inputs}
\begin{itemize}[nosep, leftmargin=*]
\item \texttt{query} (string)
\item \texttt{original\_instruction} (string)
\item \texttt{positive\_passage} = P$^{+}$ (string)
\item \texttt{specific\_instruction\_negative} = N$^{\star}$ (string)
\item \texttt{remaining\_instruction\_negatives} = N$_1$\ldots N$_k$ (array; may be empty)
\end{itemize}

\medskip
\textbf{Output}\\
You should reason step by step, and the final answer should be in the following XML format:\\[2pt]
\texttt{<answer>}\\
\quad\texttt{<new\_instruction>}[your new instruction]\texttt{</new\_instruction>}\\
\texttt{</answer>}\\[2pt]
If you think this task is too hard to achieve, you should simply return \texttt{<answer>None</answer>}.

\medskip
\textbf{Definitions}
\begin{itemize}[nosep, leftmargin=*]
\item \textit{Relevant to the pure query}: reasonably satisfies the user's intent without any extra instruction.
\item \textit{Instruction negative}: relevant to the pure query, but \textbf{excluded by the instruction} (e.g., by scope, geography, timeframe, format, source constraints, audience level).
\end{itemize}

\medskip
\textbf{Method}
\begin{enumerate}[nosep, leftmargin=*, label=\arabic*)]
\item \textit{Profile passages}
  \begin{itemize}[nosep, leftmargin=*]
  \item Identify attributes of P$^{+}$ (domain, geography, timeframe, audience, medium/format, methodology, sources, constraints).
  \item Identify attributes of N$^{\star}$ that \textbf{distinguish it from P$^{+}$}.
  \item Skim each N$_i$ to note attributes you must \textbf{keep excluded}.
  \end{itemize}
\item \textit{Choose reversal levers}\\
  Craft a new instruction that:
  \begin{itemize}[nosep, leftmargin=*]
  \item \textbf{Positively selects} N$^{\star}$'s attributes (so N$^{\star}$ becomes positive), and
  \item \textbf{Excludes} P$^{+}$ via $\geq$1 \textbf{hard, objective constraint} (so P$^{+}$ becomes an instruction negative), while
  \item \textbf{Does not inadvertently admit} any N$_i$ (keep them instruction negatives).
  \end{itemize}
  Useful levers: domain narrowing, region, timeframe/recency, audience level, style/format (e.g., ``equations only'', ``bulleted checklist''), methodology/evidence type, required artifacts (e.g., runnable code in a specific language), explicit exclusions (e.g., ``exclude major capitals'', ``exclude beach/island topics'').
\item \textit{Diversity requirement}\\
  Ensure the new instruction's \textbf{format and perspective differ} from the \texttt{original\_instruction} (e.g., switch voice, deliverable type, constraint style). Avoid trivial rewording.
\item \textit{Sanity checks}
  \begin{itemize}[nosep, leftmargin=*]
  \item Would P$^{+}$ be filtered out by the new constraints? If not, tighten them.
  \item Is N$^{\star}$ clearly included by the positive selectors? If not, pivot constraints toward N$^{\star}$'s attributes.
  \item Do all N$_i$ still get excluded? If any slip in, add explicit exclusions or tighten scope.
  \end{itemize}
\item \textit{Conciseness}\\
  The \textbf{new instruction must be one or two sentences}, imperative, concrete, and unambiguous.
\end{enumerate}

\medskip
\textbf{Guardrails}
\begin{itemize}[nosep, leftmargin=*]
\item Do \textbf{not} reference passage IDs or this meta-task; exclude or include \textbf{by attribute} only.
\item Do \textbf{not} modify the query or passages.
\end{itemize}

\medskip
\textbf{Checklist} (before you output)
\begin{itemize}[nosep, leftmargin=*]
\item P$^{+}$ becomes an instruction negative.
\item N$^{\star}$ becomes positive.
\item All N$_i$ remain instruction negatives.
\item Instruction is concise ($\leq$2 sentences) and \textbf{diverse} vs.\ the \texttt{original\_instruction}.
\item No meta-task language or passage IDs appear in the instruction.
\end{itemize}

\medskip
\textbf{Now perform the task with the following data:}
\begin{itemize}[nosep, leftmargin=*]
\item Query: \texttt{\{\{ query \}\}}
\item Original instruction: \texttt{\{\{ original\_instruction \}\}}
\item Original positive passage: \texttt{\{\{ original\_positive \}\}}
\item Specific instruction negative passage: \texttt{\{\{ specific\_instruction\_negative \}\}}
\item All remaining instruction negative passages:\\
\texttt{\{\% for negative in remaining\_negatives \%\}}\\
\quad Negative passage \texttt{\{\{ loop.index \}\}}: \texttt{\{\{ negative \}\}}\\
\texttt{\{\% endfor \%\}}
\end{itemize}
\end{tcolorbox}

\end{document}